\documentclass[showpacs,showkeys,12pt,preprint,preprintnumbers,nofootinbib,groupedaddress,superscriptaddress,amsmath,amssymb,axodraw]{revtex4}

\usepackage[dvips]{color}
\usepackage{graphicx}
\usepackage{epsfig}    
\usepackage{dcolumn}
\usepackage{bm}
\usepackage{here}

\def\Journal#1#2#3#4{{#1} {#2} (#4) #3}

\def\NPB{{Nucl. Phys.} B}
\def\PLB{{Phys. Lett.}  B}
\def\PRL{Phys. Rev. Lett.}
 
\def\PRD{{Phys. Rev.} D}

\def\EPC{{Eur. Phys. J.} C}
\def\PTP{Prog.~Theor.~Phys.}
\def\JHEP{J.~High Energy Phys.}
\def\MPLA{{Mod.~Phys.~Lett.} A}

\allowdisplaybreaks[4]

\begin{document}

\title{The $H^\pm W^\mp Z^0$ vertex and single 
charged Higgs boson production via $WZ$ fusion 
at the Large Hadron Collider }

\author{Eri Asakawa}
\email{eri@yitp.kyoto-u.ac.jp}
\affiliation{Yukawa Institute for Theoretical Physics, Kyoto
University, \\ Kyoto 606-8502, Japan}

\author{Shinya Kanemura}
\email{kanemu@het.phys.sci.osaka-u.ac.jp}
\affiliation{Department of Physics, Osaka University, 
Toyonaka, Osaka 560-0043, Japan}

\preprint{YITP-05-31}
\preprint{OU-HET 533}

\pacs{12.60.Fr, 14.80.Cp}

\keywords{Higgs, LHC, Beyond the Standard Model}

\begin{abstract}

There is a variety of new physics scenarios which deduce
extended Higgs sectors in the low energy effective theory. 
The coupling of a singly-charged Higgs boson ($H^\pm$) 
with weak gauge bosons, $H^\pm W^\mp Z^0$, directly depends on the 
global symmetry structure of the model, so that its experimental 
determination can be useful to test each scenario.
We discuss predictions on this coupling in several models, 
such as the model with additional real and complex triplets,  
the Littlest Higgs model, the two Higgs doublet model and 
the minimal supersymmetric standard model. 
In order to measure the $H^\pm W^\mp Z^0$ coupling 
we consider single $H^\pm$ production via 
the $WZ$ fusion mechanism at the CERN 
Large Hadron Collider. 
The production rate is hierarchically different among these models, 
so that this process can be useful to explore new physics scenarios. 

\end{abstract}

\maketitle

\section{Introduction}

Current precision data provide important indications 
for the structure of the electroweak symmetry breaking sector.
In particular, the experimental value of the electroweak 
rho parameter ($\rho$) is very close to unity\cite{pdg}. 
In the Standard Model (SM) 
with a scalar doublet field, this experimental requirement 
is automatically satisfied  due to the custodial $SU(2)$ symmetry, 
by predicting the rho parameter to be exactly unity at the tree level.
The data are then used to constrain the mass of the Higgs boson 
at the quantum level\cite{lepewwg},   
which is the last undetermined parameter of the model. 
It is well known that
the tree level prediction of $\rho=1$ is a common feature of 
Higgs models with only doublets (and singlets)\cite{cs,hhg}.
In Higgs models with other $SU(2)$ representations 
such as triplets, the rho parameter is generally not unity 
at the tree level unless specific combinations are assumed 
among scalar multiplets. 
Although in such models the parameters are severely constrained 
by the rho parameter data, some new physics models would give 
a motivation to study phenomenology of these 
exotic representations. 

In extended Higgs models, which would be deduced
in the low energy effective theory of new physics models,  
additional Higgs bosons like charged and CP-odd scalar bosons  
are predicted.  
Phenomenology of these extra scalar bosons strongly depends on 
the characteristics of each new physics model. 
By measuring their properties like masses, widths, 
production rates and decay branching ratios,  
the outline of physics beyond the electroweak scale 
can be experimentally determined. 
The coupling of a singly-charged Higgs boson ($H^\pm$) with the weak gauge 
bosons, $H^\pm W^\mp Z^0$, is of particular importance 
for such an approach. Its magnitude is directly related 
to the structure of the extended Higgs sector under global 
symmetries\cite{HWZ,hhg,CS-2HDM}.  
It can appear at the tree level in models with scalar triplets, 
while it is induced at the loop level in multi scalar doublet models. 

Models with scalar triplets may provide a solution for the origin 
of tiny neutrino masses. The triplet fields also appear  
in left-right symmetric models.  
The Littlest Higgs model\cite{LittlestHiggs,LH-ph} 
and some extra dimension models\cite{extraD-triplet} 
predict an additional complex triplet as well.  
The most discriminative feature of these triplet models 
is the prediction of both doubly- and singly-charged 
Higgs bosons\cite{hhg}. 
In particular, detection of doubly charged Higgs 
bosons is a clear evidence for such exotic representations.   
On the other hand, there are lots of motivations 
to consider models with multi Higgs doublets, 
such as supersymmetry, topcolor\cite{topcolor}, 
Little Higgs models\cite{little} and the model of 
gauge-Higgs unification\cite{hosotani}. 
Tiny neutrino masses ({\it e.g.} the Zee model\cite{zee}),
and extra CP violating phases\cite{CPV} which may be required for 
the realization of electroweak baryogenesis\cite{EWBG} 
also can be studied by introducing multi scalar doublets 
(plus singlets) in the electroweak scale. 
These multi doublet models predict singly-charged scalar bosons.
To distinguish these extended Higgs models at collider experiments, 
the $H^\pm W^\mp Z^0$ vertex can be an useful probe.
  
At the Fermilab Tevatron with the proton-antiproton collision 
energy of 2 TeV, $H^\pm$ may be 
predominantly produced via the gauge boson associated production.  
They are expected to be produced via the gluon-bottom 
fusion at the CERN Large Hadron Collider (LHC) 
where incident protons collide with the energy of 14 TeV, 
assuming that they couple to quarks. 
Once a charged Higgs boson is produced, we may consider 
the decay into a $W^\pm Z^0$ pair as long as it is kinematically allowed.  
The decay rate of $H^\pm \to W^\pm Z^0$ has been evaluated 
in the minimal supersymmetric standard model (MSSM) and 
the two Higgs doublet model (2HDM) 
in Refs.~\cite{HWZ-2HDM0,HWZ-2HDM1,HWZ-2HDM2}, and also in 
the models with triplet Higgs fields in Ref.~\cite{HWZ-Tevatron}. 
Impact of the $H^\pm W^\mp Z^0$ vertex 
on the physics with $e^+e^-$ collisions  
has been studied in the triplet model\cite{eeHW-TRIP}. 
Single $H^\pm$ production associated with 
a $W$ boson, $e^+e^- \to H^\pm W^\mp$, 
may also be useful to study this
vertex at a future linear 
collider\cite{eeHW-2HDM1,eeHW-2HDM2,singleH+,eeHW-MSSM}. 

In this letter, we discuss the 
$H^\pm W^\mp Z^0$ vertex in various scenarios. 
Predictions on the form factors of the vertex 
are studied in models with the two doublet 
fields and also with triplets.  
We then consider testing the $H^\pm W^\mp Z^0$ coupling  
via single $H^\pm$ production 
in the $WZ$ fusion process at the LHC.
In general, Higgs boson production by vector boson fusion has 
advantages as compared to the other production processes, 
because the signal can be reconstructed completely 
and jet production is suppressed 
in the central region due to lack of color flow 
between the initial state quarks\cite{vbf}.  
We evaluate the production rate of 
$pp \to W^{\pm\ast} Z^{0\ast} X \to H^\pm X$ 
with the effective $H^\pm W^\mp Z^0$ coupling 
in the effective vector boson approximation\cite{eff-W-approx}.   

As the reference models, 
we here consider the model with a complex doublet with 
the hypercharge $Y=1$, 
a real triplet ($Y=0$) and a complex triplet ($Y=2$)\cite{TRIP};  
the Littlest Higgs model\cite{LittlestHiggs,LH-ph,dawson} where 
the low energy effective theory includes the SM-like 
Higgs doublet with an additional complex triplet with $Y=2$;  
the general 2HDM; and the 
MSSM. These models predict different values for 
the production rate, so that each new physics scenario can be 
tested through the $WZ$ fusion process at the LHC.

\section{The vertex}
\label{Sec:vertex}

We here discuss general characteristics of the $H^\pm W^\mp Z^0$ 
coupling. The vertex (see Fig.~1) is defined as 
\begin{align}
i g m_W V_{\mu\nu} 
\epsilon_W^{\ast\mu}(p_W, \lambda_W) 
\epsilon_Z^{\ast\nu}(p_Z, \lambda_Z) , 
\end{align}
where $g$ is the weak gauge coupling, $m_W^{}$ is the mass of 
the weak boson $W^\pm$, and $\epsilon_V^{\ast\mu}(p_V, \lambda_V)$ 
($V=W$ and $Z$) are polarization vectors for 
the outgoing weak gauge bosons with the momentum $p_V^{}$ and 
the helicity $\lambda_V^{}$. The tensor $V_{\mu\nu}^{}$ is 
decomposed in terms of three form factors 
as\cite{HWZ-2HDM0,HWZ-2HDM1,HWZ-2HDM2}
\begin{align}
V_{\mu\nu} = F g_{\mu\nu}
            +\frac{G}{m_W^2}{p_Z^{}}_\mu {p_W^{}}_\nu
            +\frac{H}{m_W^2}\epsilon_{\mu\nu\rho\sigma}
             {p_Z^{}}^\rho {p_W^{}}^\sigma, \label{eq:HWZ}
\end{align}
where the antisymmetric tensor $\epsilon_{\mu\nu\rho\sigma}$ 
is defined so as to satisfy $\epsilon_{0123}^{}=-1$. 
The values of $F$, $G$ and $H$ depend on 
the detail of the model. We discuss them in
several models later.  

\begin{figure}[t]
\includegraphics[width=12cm]{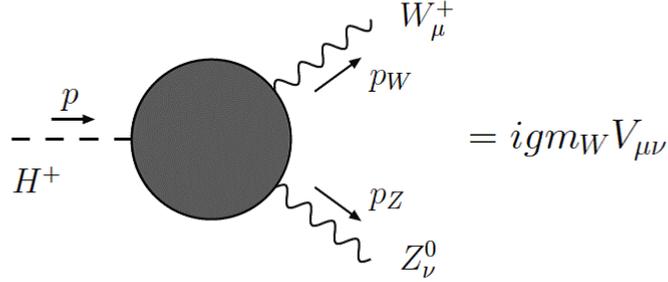}
\caption{The $H^\pm W^\mp Z^0$ vertex.}
\label{Fig:VVfusion}
\end{figure}

The three form factors $F$, $G$ and $H$ in Eq.~(\ref{eq:HWZ}) 
respectively correspond to the coefficients 
$f_{HWZ}^{}$,
$g_{HWZ}^{}$ and 
$h_{HWZ}^{}$
of three operators in the effective Lagrangian, 
\begin{align}
{\cal L_{\rm eff}} = 
   f_{HWZ}^{} H^\pm W^\mp_\mu Z^\mu 
+  g_{HWZ}^{} H^\pm F^{\mu\nu}_Z F^W_{\mu\nu}
+  h_{HWZ}^{} i \epsilon_{\mu\nu\rho\sigma} 
    H^\pm F^{\mu\nu}_Z F^W_{\rho\sigma} + H.c. , \label{eq:eff}
\end{align}
where $F^V_{\mu\nu}$ ($V=W$ and $Z$) are the field strength tensors  
for weak gauge bosons. In Eq.~(\ref{eq:eff}),  
$H^\pm W^\mp_\mu Z^\mu$ is the dimension 3 operator while 
the rest two are dimension 5, so that only $F$  
may appear at the tree level. 
Models with triplet representations can predict the nonzero 
value of $F$ at the tree level.
On the other hand, in multi Higgs doublet models, 
the vertex is induced only at the loop level because of 
the custodial $SU(2)$ symmetry in the kinetic term of
the scalar doublet fields\cite{CS-2HDM}. 
The one-loop contributions of 
the heavy particles in the loop to $f_{HWZ}^{}$, $g_{HWZ}^{}$ and 
$h_{HWZ}^{}$ are described as 
\begin{align}
   f_{HWZ}^{} \sim 
 \frac{g^2}{\cos\theta_W^{}(4\pi)^2 v}  M_i^2 \log M_i, \;\;\;\;\;\;
   g_{HWZ}^{} 
\sim 
   h_{HWZ}^{} \sim
 \frac{g^2}{\cos\theta_W^{}(4\pi)^2 v}  \log M_i,  \label{eq:pc}
\end{align}
by using the power counting,  
where $M_i$ represents the mass of the particle in the 
loop, $\theta_W^{}$ is the Weinberg angle, and $v$ ($\simeq 246$ GeV) 
is the vacuum expectation value (VEV).  
Therefore, as long as the loop-induced 
vertex is substantial due to the dynamics of the 
loop particle, $F$ gives the  
dominant contribution. This observation is correct 
for the quark loop contributions\cite{HWZ-2HDM1,HWZ-2HDM2},  
and for those of heavy Higgs bosons with the non-decoupling 
property where their masses are proportional 
to the VEV\cite{THDM-nondec,HWZ-2HDM2}.

The decay rate of $H^\pm \to W^\pm_i Z^0_i$,  
where $i=L$ represents the longitudinal polarization 
and $i=T$ does the transverse polarization, is expressed 
by
\begin{align}
\Gamma(H^\pm\to W^\pm_i Z^0_i) = 
m_{H^\pm}^{} \frac{\lambda^{1/2}(1, w, z)}{16\pi} |M_{ii}|^2, 
\end{align}
where 
$\lambda(a,b,c)=(a-b-c)^2-4 b c$, 
$w=(m_W^2/m_{H^\pm}^2)$ and  
$z=(m_Z^2/m_{H^\pm}^2)$. 
The longitudinal and transverse contributions are 
given in terms of $F$, $G$ and $H$ by 
\begin{align}
|M_{LL}|^2 &= \frac{g^2}{4 {z}}
      \left| (1-{w}-{z}) F + 
             \frac{\lambda(1, {w}, {z})}{2 {w}} G\right|^2, \\
|M_{TT}|^2 &= g^2
      \left\{ 2 {w} |F|^2 
   + \frac{\lambda(1, {w}, {z})}{2 {w}} 
            |H|^2  \right\}.
\end{align}
For the case of $m_{H^\pm} \gg m_{Z}$, we have
$|M_{TT}|^2/|M_{LL}|^2 \sim 8 m_W^2 m_Z^2/m_{H^\pm}^4$, so that 
the decay into a longitudinally polarized weak boson pair 
dominates that into a transversely polarized one.
We use these formulas for the evaluation of the production rate 
of $p p \to W^{\pm\ast} Z^{0\ast} X \to H^\pm X$ 
in the effective vector boson approximation in Sec IV.

\section{Predictions on the $H^\pm W^\mp Z^0$ vertex in various models}
\label{Sec:predictions}

\subsection{Models with triplets}

In models with triplets, the $H^\pm W^\mp Z^0$ vertex 
generally appears at the tree level. 
A common feature of the tree level contribution 
to the form factor $F$ is the fact that 
it is proportional to the VEV of the triplet field\cite{hhg,TRIP},
$F   \propto  v'/v$, 
where $v$ and $v'$ represent the VEVs of the doublet 
and the triplet in the model, respectively. 
When more than one triplet appear in the model, 
$v'$ should be taken as the combination of the VEVs for them. 

In general, models including a triplet field predict 
the value of the rho parameter not to be unity with the 
deviation proportional to $v'$, 
so that $v'$ in such models is strictly constrained 
to be much smaller than $v$; i.e., $F \ll 1$. 
For example, we consider the low energy effective theory 
of the Littlest Higgs model\cite{LittlestHiggs,LH-ph}, 
which predicts a 
complex triplet field in addition to the SM like doublet field. 
The Littlest Higgs model is the model with 
the $SU(5)$ global symmetry, with a locally gauged subgroup 
$G_1 \otimes G_2 = 
[SU(2)_1 \otimes U(1)_1] \otimes [SU(2)_2 \otimes U(1)_2]$.  
After the global $SU(5)$ symmetry breaks down to $SO(5)$ 
by the VEV of the order $f$, 
the $SU(5)$ 24 dimensional scalar field provides 14 degrees of 
freedom to the massless Goldstone bosons, 
which transform under the electroweak gauge symmetry 
as a real singlet, a real triplet, a complex doublet 
and a complex triplet. The degrees of freedom of 
the real singlet field and the real triplet are absorbed as 
the longitudinal components of the broken gauge groups. 
With the aid of the Coleman-Weinberg mechanism the remaining 
complex doublet and the complex triplet obtain 
their masses of orders $v$ and $f$, respectively, and 
trigger the electroweak symmetry breaking. 
Therefore, an additional complex triplet field appears in the 
effective theory. 
The form factor $F$ of the $H^\pm W^\mp Z^0$ coupling is given 
in this model by 
\begin{align}
  F^{\rm (LLH)} = \frac{4}{\cos\theta_W^{}} \frac{v'}{v}.
\end{align}
The electroweak data indicate 
$1 \lesssim v' \lesssim 4$ GeV for $f=2$ TeV\cite{dawson}.  
The mass $m_\Phi^{}$ of the triplet field $\Phi$,  
of which $H^\pm$ is a component, is expressed 
at the leading order by\cite{LH-ph} 
\begin{align}
  m_{\Phi}^2 = \frac{2 m_h^2 f^2}
{ v^2 \left\{ 1 - (4 v' f/v^2)^2 \right\} }, 
\end{align}
where $m_h$ is the mass of the SM-like Higgs boson.
We here consider the case with $m_h=115$ GeV,  $f=1$ TeV (2 TeV), 
$v' = 5$ GeV ($4$ GeV), and $m_{H^\pm}^{}=700$ GeV (1.56 TeV)
as a reference: i.e., 
the value of the form factor $F$ is 
$|F^{\rm (LLH)}|^2 \simeq 0.0085$ (0.0054).

In the model with additional real and complex triplet fields, 
the rho parameter can be set to be unity at the tree level, 
by imposing the custodial symmetry; i.e.,
$v_{\rm r}' = v_{\rm c}'(=v')$, 
where $v_{\rm r}'$ and $v_{\rm c}'$ are respectively 
the VEVs of the real and the complex triplet field\cite{TRIP,cg,thetaH}. 
After the electroweak symmetry breaking
the remaining degrees of freedom is a five-plet ($H_5^\pm$), 
a three-plet ($H_3^\pm$) and two singlets under the custodial symmetry.
In the case without mixing between the five-plet and three-plet, 
only the three-plet couples to fermions, while  
the singly-charged Higgs boson $H^\pm_5$ of the five-plet couples 
to $W^\mp Z^0$. The form factor is given by\cite{thetaH}  
\begin{align}
   F^{\rm (triplet)} =  \frac{1}{\cos\theta_W^{}} \sin\theta_H^{},   
\end{align}
where $\sin\theta_H^{}=\sqrt{8  v'^2/(v^2+8 v'^2)}$.
In this model, the constraint from the rho parameter 
is weak and $\tan\theta_H^{}$ can be taken to be of order 1. 
The strongest experimental bound on $v'/v$ 
comes from the $Z b \bar b$ result.  
The limits at 95\% CL are $\tan\theta_H^{} \lesssim 0.5, 1$ and $1.7$
for the mass of $H_3$ to be $0.1,0.5$ and $1$ TeV, respectively\cite{logan}.
We here take $\tan\theta_H^{}$ to be 0.5 and $m_{H_5}$,  
the mass of charged Higgs boson from the five-plet,  
to be $200$ GeV; i.e., $|F^{\rm (triplet)}|^2 \simeq 0.26$. 

\subsection{Multi Higgs doublet models}

In models with multi Higgs doublets (and singlets), 
the $H^\pm W^\mp Z^0$ coupling is forbidden at the tree level\cite{HWZ} 
due to the custodial $SU(2)$ symmetry in the kinetic term 
of the scalar doublets.  
We here discuss the cases of the 2HDM and the 
MSSM\cite{HWZ-2HDM0,HWZ-2HDM1,HWZ-2HDM2}. 
The vertex can be induced at the one-loop 
level corresponding to the terms 
\begin{align}
{\rm tr}\left[
\tau_3 (D_\mu {\cal M})^\dagger (D_\mu {\cal N}) \right], 
{\rm tr}\left[
\tau_3  {\cal M}^\dagger {\cal N}
        F^{\mu\nu}_Z F_{\mu\nu}^W \right], {\rm and }\hspace{2mm}
i \epsilon_{\mu\nu\rho\sigma} {\rm tr}\left[
\tau_3 {\cal M}^\dagger {\cal N}
        F^{\mu\nu}_Z F^{\rho\sigma}_W \right], \label{efflag}
\end{align}
in the effective Lagrangian\cite{HWZ-2HDM2}, according to the deviation 
from the custodial $SU(2)$ invariance in each part of 
the Lagrangian. In Eq.~(\ref{efflag}), 
${\cal M}$ and ${\cal N}$ are $2 \times 2$ matrices 
defined by ${\cal M}=(i \tau_2 \Phi^\ast, \Phi)$ and 
${\cal N}=(i \tau_2 \Psi^\ast, \Psi)$, where $\Phi$ and $\Psi$ 
are the two scalar doublet fields in the gauge eigenstate basis\cite{georgi}; 
i.e., $\langle \Phi \rangle = (0, v/\sqrt{2})^T$ and 
$\langle \Psi \rangle = 0$.  
Under $SU(2)_L \otimes SU(2)_R$, 
${\cal M}$ and ${\cal N}$ transform  
as ${\cal M} \to g_L {\cal M} g_R^\dagger$ and 
${\cal N} \to g_L {\cal N} g_R^\dagger$, where 
$g_{L,R} \in SU(2)_{L,R}$. 
It is clear that the terms in Eq.~(\ref{efflag}) 
are invariant under $SU(2)_L$, but not under $SU(2)_R$: 
i.e., the custodial symmetry is explicitly 
broken in these terms.

In the 2HDM, there are two sources to enhance the 
loop-induced form factors; i.e., the contribution from 
the top-bottom loop and those from the Higgs-boson loop.
The large mass difference between top and bottom quarks 
indicates a large breakdown of the custodial $SU(2)$ symmetry 
in the top-bottom quark sector, and the loop 
induced $H^\mp W^\mp Z^0$ vertex can be sizable 
with quadratic power contributions of the top quark mass.     
The leading contribution 
can be extracted from the result of the full one-loop 
calculation\cite{HWZ-2HDM1}   as
\begin{align}
F^{(t-b \;\rm loop)} \simeq  \frac{N_c}{(4\pi)^2\cos\theta_W^{}} 
\frac{m_t^2}{v^2} \cot\beta, \;\;\;\;\;\;\;\;\;\;\;\;\;
 {\rm for} \;\; m_b \ll m_t, 
\end{align}
where $\tan\beta$ is the ratio of VEVs of Higgs bosons\footnote{
If we consider the situation with $m_b \sim m_t$, 
the leading contribution is extracted for Model II 2HDM 
as\cite{HWZ-2HDM1} 
\begin{align}
F^{(t-b \;\rm loop)} 
\simeq \frac{N_c}{(4\pi)^2 \cos\theta_W^{}} \frac{m_t^2-m_b^2}{3v^2} 
(\tan\beta+\cot\beta),   \;\;\; {\rm if} \;\; m_b \sim m_t.
\end{align}
In the limit of $m_b \to m_t$, the leading contribution to  
$F$ becomes zero, according 
that the Yukawa interaction for the third generation quarks 
is invariant under $SU(2)_L \times SU(2)_R$\cite{CS-2HDM,screening}.  
This is described by expressing 
${\cal L}_{\rm Yukawa}^{\rm 3rd\; gen.} = {\bar Q}_L {\cal M}_{21}' Q_R$, 
where $Q_{L,R}=(t_{L,R}^{}, b_{L,R}^{})^T$ and 
${\cal M}_{21}' = \left[  i \tau_2 \Phi_2'^\ast, \Phi_1' \right]$ 
transform 
as $Q_{L,R} \to g_{L,R}^{} Q_{L,R}$ and  
${\cal M}_{21}' \to g_L {\cal M}_{21}' g_R^\dagger$ where     
$\Phi_{i}'= y_i \Phi_{i}$ ($i=1,2$) with $y_1$ ($y_2$) 
to be the bottom (top) Yukawa coupling and 
 $\Phi_{1,2}$ being the Higgs doublets and $\tau_i$ ($i=1-3$) 
being the Pauli matrix. }.
Notice that this expression is independent of 
the type of Yukawa interaction, either Model I or Model II\cite{hhg}.
The values of $F^{(t-b \;\rm loop)}$ are given by 
$F^{(t-b \;\rm loop)} \simeq 0.01 \cot\beta$; i.e.,
$|F^{(t-b \;\rm loop)}|^2 
\simeq 10^{-3}, 10^{-4}$ and $10^{-5}$ for $\tan\beta=0.3,1$ and $3$, 
respectively.  
The value of $\tan\beta$ is bounded from below 
by the condition that the top Yukawa coupling 
should not be too large; i.e., 
$\tan\beta > 0.2-0.3$\cite{eeHW-2HDM1}. 
The one-loop diagrams of heavy neutral Higgs bosons 
can also contribute to this vertex  
when the mass difference 
between the charged Higgs boson and the CP-odd Higgs boson is 
large\cite{HWZ-2HDM2}. 
This mass splitting implies large breaking of 
the custodial $SU(2)$ symmetry under 
${\cal M} \to g_L {\cal M} g_R^\dagger$ and 
${\cal N} \to g_L {\cal N} g_R^\dagger$ 
in the Higgs potential. 
The constraint from the rho parameter can be satisfied 
by imposing ``another'' global $SU(2)$ symmetry 
under ${\cal M}_{21} \to g_L {\cal M}_{21} g_R^\dagger$, where 
${\cal M}_{21} = \left[  i \tau_2 \Phi_2^\ast, \Phi_1 \right]$
in the Higgs potential\footnote{
The choice 
$m_{H^\pm}^{} =m_H^{}$ and $\sin(\alpha-\beta)=-1$, 
or $m_{H^\pm}^{} =m_h^{}$ and $\sin(\alpha-\beta)=0$, 
corresponds to this case, where $m_h$ and $m_H^{}$ are 
the masses of the lighter and heavier 
neutral CP-even Higgs boson, and $\alpha$ is the mixing angle 
between them.}.
The contribution of the Higgs boson loop can be 
important for $\tan\beta \gtrsim 3$, where the top-bottom 
loop contribution becomes suppressed because of the smaller 
Yukawa couplings\cite{eeHW-2HDM1}. 
However, too large mass splitting between 
the charged Higgs boson and the CP-odd Higgs boson causes 
a problem from the argument of perturbative unitarity\cite{pu,2hdm_pu}. 
Consequently, contributions from the bosonic loop 
to $F$ is constrained as 
$|F^{({\rm bosonic\; loop})}|^2 \lesssim 10^{-5}$ 
for $3 \lesssim \tan\beta \lesssim 10$. 
Therefore, as the reference value of the 2HDM, we can take the value 
$|F^{\rm (THDM)}|^2 \sim 10^{-3}$, $10^{-4}$ and  
$10^{-5}$ for $\tan\beta=0.3$, $1$ and $3-10$, respectively. 

In the MSSM, the loop effect of super partner 
particles can enhance the vertex especially 
in the moderate values of $\tan\beta$, where
the top-bottom loop contribution becomes suppressed. 
The new contributions become large according to  
the breakdown of global $SU(2)$ symmetry in 
the sfermion and chargino/neutralino sector.   
They can dominate the top-bottom loop contribution 
especially in the region of $\tan\beta \gtrsim 3$. 
However, the magnitude is at most $|F|^2 \sim 10^{-5}$\cite{eeHW-MSSM}, 
because of the decoupling property of super particles. 
On the other hand, as masses of the heavy Higgs bosons of the MSSM 
are approximately independent of the VEV and are nearly degenerate, 
the contribution from the Higgs-boson loop are small. 
Therefore, 
as a reference of the MSSM, we take 
$|F^{\rm (MSSM)}|^2 \lesssim 10^{-5}$ for $\tan\beta \gtrsim 3$.

\section{Single $H^\pm$ production via $WZ$ fusion at the LHC}
\label{Sec:WZfusion}

Let us study the impact of the $H^\pm W^\mp Z^0$ vertex 
on the production cross section of 
$pp \to W^{\pm\ast} Z^{0\ast} X \to H^\pm X$ 
in the models discussed above. The vector boson fusion 
is a pure electroweak process with high-$p_T$ jets going into 
the forward and backward directions
from the decay of the produced scalar boson without color flow 
in the central region.
The signal can be reconstructed, and the 
backgrounds are expected to be sufficiently reduced by 
appropriate kinematic cuts.

The hadronic cross section for $pp \to H^\pm X$ via 
$WZ$ fusion is expressed in the effective vector boson 
approximation\cite{eff-W-approx} by
\begin{align}
\sigma_{\rm eff}(s,m_{H^\pm}^2) \simeq  
\frac{16\pi^2}{ \lambda(1,w,z) m_{H^\pm}^3} 
\sum_{\lambda=T,L} \Gamma(H^\pm \to W^\pm_\lambda Z^0_\lambda)
       \tau \left.\frac{d {\cal L}}{d \tau} 
      \right|_{pp/W^\pm_\lambda Z^0_\lambda}  
       , 
\end{align}
where $\tau=m_{H^\pm}^{2}/s$, and 
\begin{align}
\left.\frac{d {\cal L}}{d \tau}\right|_{pp/W^\pm_\lambda Z^0_\lambda}
= \sum_{ij} \int_\tau^{1} \frac{d\tau'}{\tau'} 
\int_{\tau'}^{1} \frac{d x}{x} f_i(x) f_j(\tau'/x)  
\left.\frac{d {\cal L}}{d \xi}\right|_{q_iq_j/W^\pm_\lambda Z^0_\lambda},
\end{align}
with $\tau'=\hat{s}/s$ and $\xi=\tau/\tau'$. Here   
$f_i(x)$ is the parton structure function for the $i$-th quark, and  
\begin{eqnarray}
\left. \frac{d{\cal L}}{d\xi} \right|_{q_iq_j/W_T^\pm Z_T^0}
&=&\frac{c}{64\pi^4}
\frac{1}{\xi} \ln \left( \frac{\hat{s}}{m_W^2} \right)
\ln \left( \frac{\hat{s}}{m_Z^2} \right)
\left[ (2+\xi)^2 \ln (1/\xi)-2(1-\xi)(3+\xi)
\right],\\
\left. \frac{d{\cal L}}{d\xi} \right|_{q_iq_j/W_L^\pm Z_L^0}
&=&\frac{c}{16\pi^4}
\frac{1}{\xi}
\left[ (1+\xi) \ln (1/\xi)+2(\xi-1)
\right],
\end{eqnarray}
where $c=(v_W^{2}+a_W^{2})(v_Z^{2}+a_Z^{2})$, and 
\begin{align}
v_W = -a_W = \frac{g}{2\sqrt{2}}, \hspace*{4mm}
v_Z = \frac{g}{\cos\theta_W} \left(
\frac{T_q^3}{2} - e_q \sin^2\theta_W
\right), \hspace*{4mm}
a_Z = - \frac{g}{\cos\theta_W} \frac{T_q^3}{2},
\end{align}
with $T_q^3$ and $e_q$ to be the weak isospin and the 
electric charge for a quark $q$, respectively.

In evaluation for the cross section here, 
the contribution from the diagram with 
the effective $H^\pm W^\mp \gamma$ vertex is neglected.    
This may be justified by the fact that 
due to the $U(1)_{\rm em}$ invariance    
there is no tree-level $H^\pm W^\mp \gamma$ coupling  
and loop-induced form factors of $H^\pm W^\mp \gamma$
do not have any quadratic mass contributions of 
the particles in the loop. 
Furthermore, for the $H^\pm W^\mp Z^0$ coupling 
the loop induced $F$ can be much greater than 
the loop induced $G$ and $H$ in the multi doublet 
models: see Eq.~(\ref{eq:pc}).
Consequently, the prediction for the cross section 
in each scenario can be obtained as 
$\sigma^{\rm (model)} \simeq |F^{\rm (model)}|^2 \times
\sigma^{|F|^2=1}$
in a good approximation.

\begin{figure}[t]
\includegraphics[width=9cm]{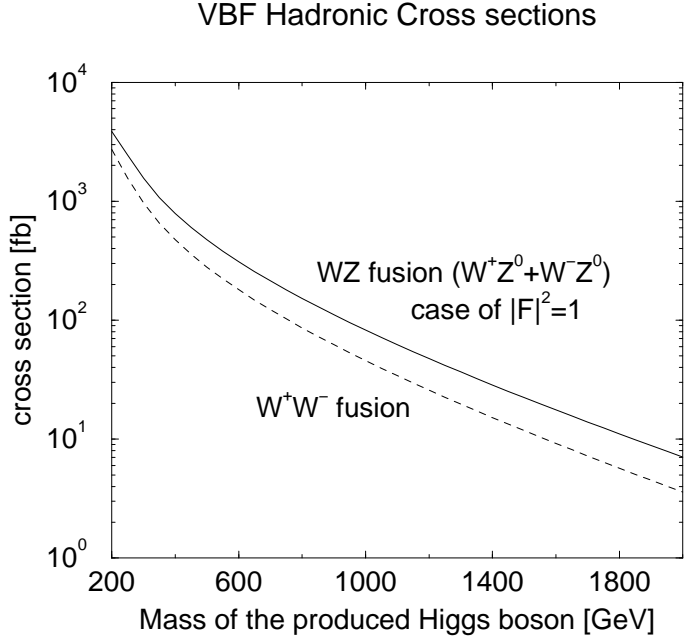}
\caption{The hadronic cross section of 
the $W^\pm Z^0$ fusion process and 
the $W^+W^-$ fusion process as a function of 
the mass of the charged and neutral Higgs bosons, respectively. 
For the $W^\pm Z^0$ fusion, the form factor $F$ is set to be unity.
The SM prediction is shown for the $W^+W^-$ fusion.
}
\label{Fig:VVfusion}
\end{figure}

In Fig.~2, the hadronic cross section of 
$pp \to W^{\pm\ast} Z^{0\ast} X \to H^\pm X$ at the LHC 
($\sqrt{s}=14$ TeV)
is shown as a function of $m_{H^\pm}^{}$. 
The form factors $F$, $G$ and $H$ of the $H^\pm W^\mp Z^0$ 
vertex are set to be 1, 0 and 0, respectively. 
The hadronic cross section of SM Higgs boson 
production via $W^+W^-$ fusion is also shown  for comparison. 
CTEQ6L is used for the parton distribution function\cite{cteq}.
Because of a $pp$ collider, the hadronic cross 
section for the $W^+ Z^0$ fusion is about $1.5-2$ times 
greater than that for the $W^- Z^0$ fusion. 
The magnitude of $\sigma^{|F|^2=1}$ can be about 
$0.4 \times 10^{4}$ fb for $m_{H^\pm}^{}=200$ GeV. 
It decreases as $m_{H^\pm}^{}$ grows, and becomes about $11$ fb 
for $m_{H^\pm}^{}=1.8$ TeV.    
If we assume that $\sigma^{\rm (model)}=1$ fb is 
a sufficient number to detect the signal, the required 
values of $|F^{\rm (model)}|^2$ are about  
$2.5 \times 10^{-4}$, 
$1.3 \times 10^{-2}$ and $10^{-1}$ 
for $m_{H^\pm}=200$ GeV, $1$ TeV and $1.8$ TeV, respectively.
In this case, 300 of $H^\pm$ 
are produced when $\sigma^{\rm (model)}=1$ fb
at the LHC with the luminosity of $300$ fb$^{-1}$.

In the model with additional real and complex triplets, 
$|F^{\rm (triplet)}|^2$ can be of order 1. 
When $|F^{\rm (triplet)}|^2 \simeq 0.26$, 
the cross sections are of order $1000$ fb and $80$ fb for 
$m_{H^\pm}=200$ GeV and $600$ GeV, respectively. 
In the Littlest Higgs model where the form factor $F$ is given 
by $|F^{\rm LLH}|^2 \sim 0.0085$ ($0.0054$) for $f=1$ TeV ($2$ TeV) 
with $m_{H^\pm}=700$ GeV ($1.56$ TeV) the cross section can 
be about 2 fb (0.1 fb). 
In the 2HDM, the one-loop induced cross section 
can be of order $4, 0.4$ and $0.04$ fb at 
$m_{H^\pm}=200$ GeV, according to the values 
$|F^{\rm (2HDM)}|^2 \sim 10^{-3}$, $10^{-4}$ and $10^{-5}$
for $\tan\beta=0.3, 1$ and $3$, respectively. 
In the MSSM, the values of $\tan\beta$ is bounded from 
below as $\tan\beta > 3-4$ by the LEP direct 
search result of the lightest Higgs boson\cite{lepewwg}.  
As $|F^{\rm (MSSM)}|^2 \lesssim 10^{-5}$ for $\tan\beta \gtrsim 3$, 
the cross section is 
at most 0.04 fb for $m_{H^\pm}^{} > 200$ GeV. 

The decay pattern of $H^\pm$ depends on the model.  
In the Littlest Higgs model in which $H^\pm$ 
couples to the top and bottom 
quarks, the main decay mode is expected to be a $t b$ pair\cite{LH-ph}. 
In the 2HDM and the MSSM, although there are potentially many decay
modes, the main mode would also be the decay into 
a $t b$ pair as long as it is kinematically allowed.  
In these cases, 
the signal would be $\ell \nu b \bar b$ ($\ell=e$ and $\mu$). 
If the cross section is 1 fb, about 60 of the signal 
events are produced at the LHC 
with the luminosity of $300$ fb$^{-1}$. 
The process can be completely reconstructed by using 
the information of the missing $p_T^{}$ and of the mass of $H^\pm$.
The main backgrounds would come from $t \bar t$, $W+j$, $WW$ and $WZ$, 
and their cross sections can be of order 1-10 pb. 
It is expected that appropriate kinematic cuts 
can reduce the backgrounds by 3-4 orders of magnitude by virtue 
of the distinct kinematic nature of vector boson fusion processes.
By assuming a good efficiency ($\sim 0.25$) 
for the double $b$ tagging, the signal events 
would be detectable. 
On the other hand, in models with triplets that  
do not couple to fermions, it would mainly decay into a $WZ$ pair. 
The model with a real and a complex triplets 
can correspond to this case. 
The signal event would be $\ell \ell \ell \nu$. 
For $\sigma^{\rm (triplet)} \simeq 100$~fb, 
about 420 of the signal events are produced, assuming 
the luminosity of 300 fb$^{-1}$. 
Again, the process can be completely reconstructed.
The main backgrounds would be 
$t \bar t$, $W+j$ and $WZ$ 
in addition to the Drell-Yan process.
We can expect that the backgrounds can be well 
rejected by the kinematic cuts, and that 
the signal can be detected\footnote{
A feasibility study for this mode 
can be seen in the context of the Higgsless model 
in Ref.~\cite{higgsless}.}.
Needless to say that in either case, the feasibility study  
has to be performed by the realistic Monte Carlo simulation.
This will be presented elsewhere\cite{WZsimulation}.  

\section{Conclusions}
\label{Sec:conclusions}

We have discussed the $H^\pm W^\mp Z^0$ vertex  
in various physics scenarios.
The magnitude of the $H^\pm W^\mp Z^0$ coupling 
directly depends on the global symmetry structure of the model, 
so that the experimental determination of the magnitude of 
the $H^\pm W^\mp Z^0$ coupling can be useful to test each scenario.
The possibility of its measurement via single charged 
Higgs boson production by $WZ$ fusion at the LHC 
has been discussed.

We have studied predictions on the $H^\pm W^\mp Z^0$ 
coupling in 
the model with additional real and complex triplets; 
the littlest Higgs model; 
the 2HDM; and 
the MSSM.  
These models predict hierarchical values 
for the form factor $F^{\rm (model)}$.   
The cross section of $pp \to W^{\pm\ast} Z^0 X \to H^\pm X$ 
has been evaluated in terms of the effective $H^\pm W^\mp Z^0$ 
coupling in the effective vector boson approximation.
In the models discussed in this letter 
(except for the MSSM), the cross section can 
exceed 1 fb: i.e., 300 of $H^\pm$ 
can be produced at the LHC for the luminosity of $300$ fb$^{-1}$.
By measuring this process we can 
obtain useful information to determine the structure of 
the Higgs sector,     
incorporating with a search for doubly charged Higgs bosons
and that for single charged Higgs bosons via the other processes.

We have shortly discussed the signal for 
the cases where $H^\pm \to tb$ is dominant 
and where $H^\pm \to W^\pm Z^0$ is dominant.
For the both cases, the backgrounds are expected 
to be considerably reduced because of the kinematic 
advantages in vector boson fusion.
The more detailed study with the 
Monte Carlo simulation is in preparation. 

\vspace{1cm}
\noindent
{\large \it Acknowledgments}

The authors would like to thank Tomio Kobayashi 
for valuable discussions and comments, 
Kaoru Hagiwara, Mihoko Nojiri, 
Yasuhiro Okada, Eibun Senaha and 
Mayumi Aoki for useful discussions.  
A part of this work started in the discussion 
during the workshop ``Physics in LHC era'' at YITP, 13-15 
December 2004 (YITP-W-04-20). 
S.K. was supported, in part, by Grants-in-Aid of the Ministry 
of Education, Culture, Sports, Science and Technology, Government of 
Japan, Grant No. 17043008.


\end{document}